\documentclass[aps,onecolumn]{revtex4}
\usepackage{amsfonts}
\usepackage{amsmath}
\usepackage{amsopn}
\usepackage{amssymb}
\usepackage{graphicx}
\usepackage{graphics}
\usepackage{color}
\usepackage{verbatim}

\begin{document}

\title{Multidimensional solitons: Well-established results and novel findings%
}
\author{Boris A. Malomed}
\affiliation{Department of Physical Electronics, School of Electrical Engineering,
Faculty of Engineering, Tel Aviv University, Tel Aviv 69978, Israel}

\begin{abstract}
A brief review is given of some well-known and some very recent results
obtained in studies of two- and three-dimensional (2D and 3D) solitons. Both
zero-vorticity (fundamental) solitons and ones carrying vorticity $S=1$ are
considered. Physical realizations of multidimensional solitons in atomic
Bose-Einstein condensates (BECs) and nonlinear optics are briefly discussed
too. Unlike 1D solitons, which are typically stable, 2D and 3D ones are
vulnerable to the instability induced by the occurrence of the critical and
supercritical collapse, respectively, in the same 2D and 3D models that give
rise to the solitons. Vortex solitons are subject to a still stronger
splitting instability. For this reason, a central problem is looking for
physical settings in which 2D and 3D solitons may be stabilized. The review
addresses one well-established topic, \textit{viz}., the stabilization of
the 3D and 2D states, with $S=0$ and $1$, trapped in harmonic-oscillator
(HO) potentials, and another topic which was developed very recently: the
stabilization of 2D and 3D free-space solitons, which mix components with $%
S=0$ and $\pm 1$ (\textit{semi-vortices} and \textit{mixed modes}), in a
binary system with the spin-orbit coupling (SOC) between its components. In
both cases, the generic situations are drastically different in the 2D and
3D geometries. In the former case, the stabilization mechanism creates a
stable ground state (GS, which was absent without it), whose norm falls
below the threshold value at which the critical collapse sets in. In the 3D
geometry, the supercritical collapse does not allow to create a GS, but
metastable solitons can be constructed.
\end{abstract}

\maketitle

\textbf{List of acronyms}: GPE -- Gross-Pitaevskii equation; GS -- ground
state; HO -- harmonic-oscillator (potential); MM -- mixed mode; NLSE --
nonlinear Schr\"{o}dinger equation; SOC -- spin-orbit coupling; SV --
semi-vortex; VA -- variational approximation; VK -- Vakhitov-Kolokolov
(stability criterion)

\section{Introduction}

Self-trapping of multidimensional (two- and three-dimensional, 2D and 3D)
localized modes in nonlinear dispersive/diffractive media, which are usually
categorized as solitons, has been a topic of great interest in many areas of
physics, as can be seen from reviews \cite{general-reviews}-\cite{RMP}. In
particular, this area of the theoretical and experimental studies is highly
relevant to nonlinear optics, as well as to nonlinear dynamics of matter
waves in Bose-Einstein condensates (BECs). Other essential realizations of
multidimensional solitons are known in models of ferromagnets \cite{ferro},
superconductors \cite{super}, and semiconductors \cite{semi}, as well as in
nuclear physics \cite{nuclear}, the classical-field theory \cite%
{Hopfion,field-theory} (including famous skyrmions \cite{skyrmion}, which
may be considered as 3D solitons), and other fields.

In addition to the their significance to fundamental studies, 2D and 3D
solitons offer potential applications. In particular, spatiotemporal optical
solitons, alias ``light bullets" \cite{Silb}, may be used as
data carriers in all-optical data-processing schemes \cite{Wagner}, and 3D
matter-wave solitons are expected to provide a basis for highly precise
interferometry \cite{interferometry}.

Unlike 1D solitons, which are usually stable objects \cite{KA,Michel},
stability is a major issue for their 2D and 3D counterparts. Indeed, the
most common cubic self-focusing nonlinearity, which can easily support 2D
and 3D formal solitons solutions, gives rise to the wave collapse (\textit{%
critical}\ \textit{collapse} in 2D and \textit{supercritical collapse} in
3D) \cite{collapse}, which completely destabilizes the respective soliton
families. In particular, the first ever example of solitons which was
introduced theoretically in nonlinear optics, \textit{viz}., the \textit{%
Townes solitons} \cite{Townes}, i.e., 2D self-trapped modes supported by the
cubic self-focusing nonlinearity (in fact, these solutions were discovered
before term ``soliton" had been coined) are subject to the
instability caused by the critical collapse, therefore they have never been
observed in the experiment. Multidimensional solitons with embedded
vorticity, alias 2D vortex rings and 3D vortex tori, which also exist in
cubic media (in particular, there are higher-order versions of Townes
solitons with embedded vorticity \cite{Minsk}), are vulnerable to a still
stronger splitting instability initiated by modulational perturbations in
the azimuthal direction; usually, the modulational instability splits the
vortex ring into a few fragments, which resemble fundamental
(zero-vorticity) solitons and suffer the destruction via the intrinsic
collapse or decay into small-amplitude waves (``radiation")
\cite{general-reviews}.

Thus, the stabilization of multidimensional fundamental and vortex solitons
in physically relevant models is an issue of great significance to
theoretical studies, and creation of the solitons in the respective physical
settings is a challenge to the experiment. This paper offers a short review
of the broad area of theoretical investigations of 2D and 3D solitons,
selecting two topics for a relatively detailed presentation, one
well-established, and one very recent. The former one, considered in Section
II, relies upon the stabilization of both fundamental (zero-vorticity)
solitons and ones with embedded vorticity $S=1$ by an external
harmonic-oscillator (HO) trapping potential, axially or spherically
symmetric one. This topic has been elaborated in detail in the course of the
past 20 years \cite{2D,Sadhan,Ueda,Dum2D,Dum3D}. A recently developed theme,
presented in Section III, is the creation of stable solitons in the 2D and
3D free space (without the help from any external potential), which mix
zero-vorticity and vortical components, by means of linear terms that
account for the spin-orbit coupling (SOC) in binary BEC. This possibility
for 2D and 3D solitons was discovered in Refs. \cite{we} and \cite{HP},
respectively.

Any stabilization mechanism must secure the solitons against the critical
collapse in 2D settings, and against the supercritical collapse in the 3D
geometry. This objective is achieved differently in 2D and 3D models, as
explained below. In the former case, the stabilizing factor breaks the
specific scale invariance, which underlies the critical collapse, and
transforms any originally unstable fundamental soliton in the
single-component model, or a mixed (fundamental-vortical) soliton in the
binary SOC system, into a ground state (GS), which does not exists in the
unstabilized 2D model. This is possible because the critical collapse sets
in when the norm of the mode exceeds a certain threshold value, and the
solitons get stabilized by pushing their norm below the collapse threshold.
Vortex solitons with $S=1$ in the 2D single-component model may be
stabilized too, as excited states, rather than the GS, provided that their
norms are not too large either. On the other hand, in the 3D setting the
threshold for the onset of the supercritical collapse is absent, hence a
true GS cannot be created. Nevertheless, the stabilization of 3D fundamental
solitons and vortical ones with $S=1$ in the single-component model, and of
3D mixed modes in the SOC system, is possible in the form of \emph{%
metastable states}, which are robust against small perturbations.

\section{Multidimensional solitons in trapping potentials}

\subsection{The basic model}

A generic method for the stabilization of both fundamental and vortical
solitons is the use of trapping potentials, the corresponding model being
based on the following nonlinear Schr\"{o}dinger equation (NLSE), alias the
Gross-Pitaevskii equation (GPE) \cite{Pit}, for the atomic mean-field wave
function $\psi \left( \mathbf{r},t\right) $, where $\mathbf{r}=(x,y,z)$ (or $%
\mathbf{r}=(x,y)$ in the 2D geometry) and $t$ are appropriately scaled
spatial coordinates and time:%
\begin{equation}
i\frac{\partial \psi }{\partial t}=-\frac{1}{2}\nabla ^{2}\psi +U\left(
\mathbf{r}\right) \psi -|\psi |^{2}\psi .  \label{GPE}
\end{equation}%
Here, $\nabla ^{2}$ is the 3D or 2D Laplacian, $U\left( \mathbf{r}\right) $
is the trapping potential, and the negative sign in front of the cubic term
implies that the nonlinearity is self-attractive.\ The trapping potential in
the 3D setting may be quasi-two-dimensional, i.e., $U=U\left( x,y\right) $
\cite{low-dim}.

In the application to optics, the evolution variable in Eq. (\ref{GPE}) is
replaced by the propagation distance ($z$), while the longitudinal
coordinate is replaced by the local time, $\tau \equiv t-z/V_{\mathrm{gr}}$,
where $V_{\mathrm{gr}}$ is the group velocity of the electromagnetic carrier
waves \cite{KA}. In this case, the trapping potential, which defines a
waveguiding channel in the bulk medium by means of the local modulation of
the refractive index, $\delta n=-U\left( x,y\right) $, may be only
quasi-two-dimensional, as it is virtually important to impose modulation of
the refractive index providing the trapping in the temporal direction. The
2D version of Eq. (\ref{GPE}), with the propagation distance $z$ replacing $t
$ , models the evolution of spatial light beams in the bulk medium with
transverse coordinates $\left( x,y\right) $.

Axially symmetric potentials (including spherically isotropic ones), $%
U=U\left( \rho ,z\right) $, where $\left( \rho ,z,\theta \right) $ is the
set of cylindrical coordinates, admit looking for solutions to Eq.~(\ref{GPE}%
) in the form of vortex modes,%
\begin{equation}
\psi =\exp \left( -i\mu t+iS\theta \right) R(\rho ,z),  \label{vort}
\end{equation}%
with real chemical potential $\mu $ and integer vorticity $S$. Real
amplitude function $R\left( \rho ,z\right) $ satisfies the stationary
equation,%
\begin{equation}
\mu u=-\frac{1}{2}\left( \frac{\partial ^{2}}{\partial \rho ^{2}}+\frac{1}{%
\rho }\frac{\partial }{\partial \rho }+\frac{\partial ^{2}}{\partial z^{2}}-%
\frac{S^{2}}{\rho ^{2}}\right) R+U\left( \rho ,z\right) R-R^{3},  \label{u}
\end{equation}%
supplemented by the boundary conditions (b.c.), $R(\rho ,z)\rightarrow 0$ at
$\rho \rightarrow \infty $ and $|z|\rightarrow \infty $, and $R\sim \rho
^{S} $ at $\rho \rightarrow 0$ (assuming $S\geq 0$).

Relevant to the experiments with BEC is the HO trapping potential \cite{Pit},%
\begin{equation}
U\left( x,y,z\right) =\frac{1}{2}\left( x^{2}+y^{2}+\Omega ^{2}z^{2}\right) ,
\label{HO}
\end{equation}%
where $\Omega ^{2}$ accounts for the anisotropy of the trap, determining the
\textit{aspect ratio}, $\sqrt{\Omega }$, between trapping lengths in plane $%
(x,y)$ and along axis $z$. Accordingly, the cases of $\Omega ^{2}\gg 1$ and $%
\Omega ^{2}\ll 1$ corresponding, respectively, to nearly two-dimensional
(``pancake-shaped", see Ref. \cite{pancake} and references
therein) and nearly one-dimensional (``cigar-shaped" \cite%
{Randy-NJP}) condensates. Equation (\ref{GPE}) conserves the energy,
\begin{equation}
E=\frac{1}{2}\int \int \int \left[ \left( \left\vert \frac{\partial \psi }{%
\partial x}\right\vert ^{2}+\left\vert \frac{\partial \psi }{\partial y}%
\right\vert ^{2}+\left\vert \frac{\partial \psi }{\partial z}\right\vert
^{2}\right) +(x^{2}+y^{2}+\Omega ^{2}z^{2})|\psi |^{2}-|\psi |^{4}\right]
dxdydz,  \label{E}
\end{equation}%
norm (scaled number of atoms in the BEC), or the total energy in optics),
\begin{equation}
N=\int \int \int \left\vert \psi (x,y,z)\right\vert ^{2}dxdydz,  \label{N}
\end{equation}%
and $z$-component of the angular momentum,
\begin{equation}
M_{z}=i\int \int \int \left( y\frac{\partial \psi }{\partial x}-x\frac{%
\partial \psi }{\partial y}\right) \psi ^{\ast }dxdydz,  \label{M}
\end{equation}%
with $\ast $ standing for the complex conjugate (in the case of the
isotropic trap, $\Omega =1$, all the three components of the angular
momentum are conserved). For the stationary state (\ref{vort}) with a
definite value of $S$, relation $M_{z}=SN$ holds.

In this section, the basic results for the structure and stability of
trapped solitons and solitary vortices are summarized for the HO potential (%
\ref{HO}), following, chiefly, Refs. \cite{Dum2D} and \cite{Dum3D}. On the
other hand, spatially periodic (lattice) potentials also offer an efficient
tool for the stabilization of 2D and 3D fundamental and vortex solitons, as
was predicted in various settings \cite{lattice,low-dim}, and demonstrated
experimentally for 2D optical solitons with embedded vorticity in photonic
lattices \cite{photorefr}, and for 2D plasmon-polariton solitons in
microcavities with a lattice structure \cite{30}.

\subsection{Stationary 3D modes: analytical and numerical results}

Before proceeding to the presentation of numerical findings, it is relevant
to display approximate analytical results which are available in the present
setting. In the linear limit, Eq. (\ref{GPE}) is tantamount to the
quantum-mechanical Schr\"{o}dinger equation for the 3D anisotropic HO. In
the Cartesian coordinates, the corresponding eigenfunctions are built as
\begin{equation}
\psi _{jkl}(x,y,z,t)=e^{-i\mu _{0}t}\Phi _{j}(x)\Phi _{k}(y)\Phi _{l}\left(
\sqrt{\Omega }z\right) ,  \label{xyz}
\end{equation}%
where $\Phi _{j}$, $\Phi _{k}$ and $\Phi _{l}$ are stationary wave functions
of 1D harmonic oscillators with quantum numbers $j,k,l$, which correspond to
energy eigenvalues $j+1/2$, $k+1/2$ and $\left( l+1/2\right) \Omega $,
respectively, the full chemical potential being
\begin{equation}
\mu _{0}=j+k+1+\left( l+1/2\right) \Omega ~.  \label{linear}
\end{equation}%
The states which go over into ones (\ref{vort}) with vorticity $S$ in the
nonlinear model are constructed as combinations of factorized wave functions
(\ref{xyz}) with $l=0$ and $j+k=S$. Since the correction to $\mu $ from the
self-attractive nonlinearity is negative, this restriction and Eq. (\ref%
{linear}) with $l=0$ impose a bound on $\mu $,
\begin{equation}
\mu \leq \mu _{0}=S+1+(1/2)\Omega .  \label{bound}
\end{equation}%
In particular, the eigenfunctions of the linear model, with vorticities $S=1$
and $S=2$, are
\begin{equation}
\psi _{\mathrm{linear}}^{(S=1)}=\psi _{100}+i\psi _{010}\equiv \rho \exp %
\left[ -\left( 2+\Omega /2\right) it+i\theta -\left( \rho ^{2}+z^{2}\right)
/2\right] ,
\end{equation}%
\begin{equation}
\psi _{\mathrm{linear}}^{(S=2)}=\psi _{200}-\psi _{020}+2i\psi _{110}\equiv
\rho ^{2}\exp \left[ -\left( 3+\Omega /2\right) it+2i\theta -\left( \rho
^{2}+z^{2}\right) /2\right] .
\end{equation}

To apply the variational approximation (VA) \cite{Progress} to the nonlinear
model, note that Eq. (\ref{u}) can be derived from Lagrangian
\begin{equation}
L=\int_{0}^{\infty }\rho d\rho \int_{0}^{+\infty }dz\left[ \left( \frac{%
\partial R}{\partial z}\right) ^{2}+\left( \frac{\partial R}{\partial \rho }%
\right) ^{2}+\left( \frac{S}{\rho }\right) ^{2}R^{2}-2\mu R^{2}+\left( \rho
^{2}+\Omega ^{2}z^{2}\right) R^{2}-R^{4}\right] .  \label{L}
\end{equation}%
A natural \textit{ansatz} for the trapped vortices is%
\begin{equation}
R(\rho ,z)=A\rho _{0}^{S}\exp \left[ -\rho ^{2}/\left( 2\rho _{0}^{2}\right)
-z^{2}/\left( 2z_{0}^{2}\right) \right] ,  \label{ans}
\end{equation}%
where amplitude $A$, as well as the radial and vertical widths, $\rho _{0}$
and $z_{0}$, are free parameters, norm (\ref{N}) of this ansatz being $%
N=\left( \pi ^{3/2}S!\right) M,M\equiv A^{2}\rho _{0}^{4}h_{0}$ (below, $M$
replaces $A$ as one of variational parameters). The trapped modes with $%
S\geq 1$ are shaped as vortex tori (``doughnuts"), see a
typical example below in Fig. \ref{fig4}. This shape is (qualitatively)
correctly captured by ansatz (\ref{ans}).

Further results produced by the VA are given here for $S=0$ and $1$, as all
vortices with $S\geq 2$ are unstable, see below. The substitution of ansatz (%
\ref{ans}) with $S=1$ in Lagrangian (\ref{L}) yields $L=\left( \sqrt{\pi }%
/8\right) M\left[ -4\mu +z_{0}^{-2}+4\rho _{0}^{-2}+4\rho _{0}^{2}+\Omega
^{2}z_{0}^{2}-M\left( 2\sqrt{2}\rho _{0}^{2}z_{0}\right) ^{-1}\right] $. The
first variational equation, $\partial L/\partial \rho _{0}=0$, yields $M=8%
\sqrt{2}\left( 1-\rho _{0}^{4}\right) h_{0}$, which predicts that the vortex
may only exist if it is narrow enough in the radial direction, $\rho _{0}<1$%
. Two other equations, $\partial L/\partial z_{0}=\partial L/\partial M=0$,
yield $z_{0}^{2}\left[ 1-2\left( 1-\mu \right) \rho _{0}^{2}-3\rho _{0}^{4}%
\right] =2\rho _{0}^{2}$, $\Omega ^{2}\rho _{0}^{2}z_{0}^{4}-2\rho
_{0}^{4}z_{0}^{2}-\rho _{0}^{2}+2z_{0}^{2}=0$.

\begin{figure}[t]
\begin{center}
\includegraphics[width=10cm]{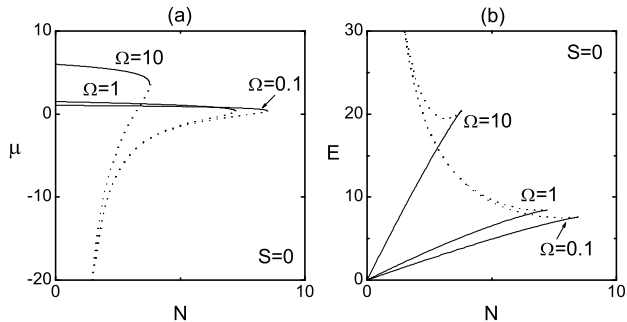}
\end{center}
\caption{(a) The chemical potential and (b) energy, defined as per Eq. (%
\protect\ref{E}), versus norm $N$ for the fundamental ($S=0$) modes trapped
in anisotropic HO potential (\protect\ref{HO}) with $\Omega =10,~1$, and $%
0.1 $, as found from numerical solution of Eq. (\protect\ref{u}). Solid and
dotted portions of the curves denote, respectively, stable and unstable
parts of the solution families, as identified by a numerical solution of the
eigenvalue problem based on Bogoliubov - de Gennes equations (\protect\ref%
{growth}). See further details in Ref. \protect\cite{Dum3D}.}
\label{fig1}
\end{figure}

\begin{figure}[t]
\begin{center}
\includegraphics[width=10cm]{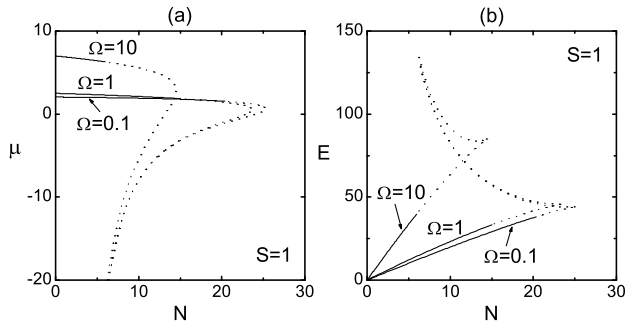}
\end{center}
\caption{The same as in Fig. \ref{fig1}, but for trapped vortex modes
with $S=1$.}
\label{fig2}
\end{figure}

Numerical solution of the variational equations generates $\mu (N)$
dependences for families of the fundamental ($S=0$) and vortical ($S=1$)
trapped modes which are displayed, severally, in Figs. \ref{fig1}(a) and \ref%
{fig2}(a) for $\Omega =10$, $1$, and $0.1$, which correspond to the
pancake-like, isotropic, and cigar-shaped configurations, respectively,
while the corresponding $E(N)$ curves are displayed in Figs \ref{fig1}(b)
and \ref{fig2}(b). The same dependences, obtained from numerical solution of
stationary equation (\ref{u}), are displayed too, demonstrating the accuracy
of the VA. It is seen that the properties of the modes trapped in the
isotropic ($\Omega =1$) and cigar-shaped ($\Omega =0.1$) HO potentials are
quite close, being strongly different from those is the pancake trap ($%
\Omega =10$). Another obvious feature of the figure is the presence of a
largest norm, which corresponds to the turning point of the $\mu (N)$ curve
and bounds the existence of the trapped modes. This limitation is caused by
the presence of the supercritical collapse in the 3D model \cite{collapse}:
if the norm is too large, the trend to the collapse, driven by the cubic
self-attraction, cannot be balanced by the quantum pressure (the gradient
part of energy (\ref{E})). The largest norm in this model was numerically
found, as a function $\Omega $, in Ref. \cite{Sadhan}. Another implication
of the possibility of the supercritical collapse is that all the stable
modes found in the 3D model are actually metastable ones. They cannot play
the role of the GS, which does not exist in the model admitting the
supercritical collapse.

\subsection{Stability of trapped 3D fundamental and vortex modes}

\begin{figure}[t]
\begin{center}
\includegraphics[width=10cm]{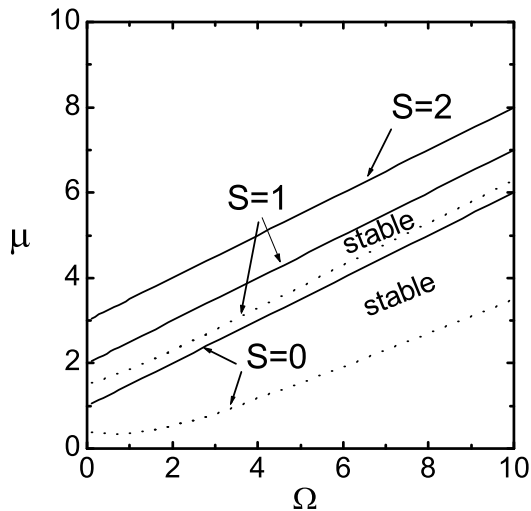}
\end{center}
\caption{Stability regions for the 3D\ trapped states with vorticities $S=0$%
, $1,$ and $2$, as per Ref. \protect\cite{Dum3D}. The solutions exist and are
stable between the upper border (solid line), which is given by $\protect\mu %
=\protect\mu _{0}$ in Eq. (\protect\ref{bound}), and the numerically found
lower stability border (dotted curve). Vortices with $S=2$ have no stability
domain.}
\label{fig3}
\end{figure}

The stability of stationary solutions produced by Eq. (\ref{u}) is
identified through the computation of (generally, complex) eigenvalues $%
\lambda $ of small perturbations. To this end, a perturbed solution to Eq. (%
\ref{GPE}) is looked for as
\begin{equation}
\psi (x,y,z,t)=[R(\rho ,z)+u(\rho ,z)\exp (\lambda t+iL\theta )+v^{\ast
}(\rho ,z)\exp (\lambda ^{\ast }t-iL\theta )]\exp \left( iS\theta -i\mu
t\right) ,  \label{pert}
\end{equation}%
where $\left( u,v\right) $ are eigenmodes of infinitesimal perturbations
corresponding to integer values of azimuthal index $L$. The substitution of
ansatz (\ref{pert}) in Eq. (\ref{GPE}) and linearization lead to\ the
\textit{Bogoliubov - de Gennes equations} \cite{Pit},%
\begin{eqnarray}
\left( i\lambda +\mu \right) u+\frac{1}{2}\left[ \frac{\partial ^{2}}{%
\partial \rho ^{2}}+\frac{1}{\rho }\frac{\partial }{\partial \rho }+\frac{%
\partial ^{2}}{\partial z^{2}}-\frac{(S+L)^{2}}{\rho ^{2}}u-\rho ^{2}\right]
u+R^{2}(v+2u) &=&0,  \notag \\
\left( -i\lambda +\mu \right) v+\frac{1}{2}\left[ \frac{\partial ^{2}}{%
\partial \rho ^{2}}+\frac{1}{\rho }\frac{\partial }{\partial \rho }+\frac{%
\partial ^{2}}{\partial z^{2}}-\frac{(S-L)^{2}}{\rho ^{2}}v-\rho ^{2}\right]
v+R^{2}(u+2v) &=&0,  \label{growth}
\end{eqnarray}%
supplemented by the boundary conditions demanding that $u(\rho ,z)$ and $%
v(\rho ,z)$ decay exponentially at $\rho \rightarrow \infty $ and $%
\left\vert z\right\vert \rightarrow \infty $, and decay as $\rho
^{\left\vert S\pm L\right\vert }$ at $\rho \rightarrow 0$. The stability of
the underlying stationary mode is secured by the condition that all
eigenvalues $\lambda $ produced by numerical solution of Eq. (\ref{growth})
\cite{Ueda} must be pure imaginary. The findings are included in Figs. \ref%
{fig1} and \ref{fig2}, where stable and unstable portions of the solution
families are distinguished. The results of the stability analysis are
collected in the diagram which displays stability regions in the plane of $%
(\mu ,\Omega )$ in Fig. \ref{fig3}.

The stability of the fundamental trapped states ($S=0$) precisely complies
with the \textit{Vakhitov-Kolokolov} (VK) criterion \cite{VK,collapse}, $%
d\mu /dN\leq 0$, which is a necessary but, in the general case, not
sufficient stability condition. On the other hand, the stability region of
the vortex mode with $S=1$ is essentially smaller than formally predicted by
this criterion. In the region with the VK criterion still holds for the
vortices, but they are actually unstable, they are vulnerable to the
above-mentioned splitting instability induced by perturbations with $L=2$
and $3$ in Eq. (\ref{pert}). The vortex modes with $S\geq 2$ were found to
be completely unstable.

A noteworthy feature revealed by Figs. \ref{fig1} and \ref{fig2} is that the
smallest and largest maximum values of $N$, up to which the trapped modes
with $S=0$ or $S=1$ remain stable, are achieved, respectively, at the
largest and smallest values of $\Omega $. In particular, the quasi-1D
``tubular" vortical modes with $S=1$ \cite{Luca}, which are
generated by the model at $\Omega \ll 1$, are tightly confined in the
corresponding cigar-shaped trap, that suppresses their destabilization by
the splitting.

As said above, at $\Omega \rightarrow \infty $ the 3D setting shrinks into
its 2D counterpart, where trapped modes with $S=1$ have a finite stability
region, $\mu _{0}-\mu _{\mathrm{cr}}\approx 0.724$, see below. Therefore,
taking into regard that Eq. (\ref{bound}) yields $\mu _{0}(\Omega )=2+\Omega
/2$ for $S=1$, the lower stability border ($\mu =\mu _{\mathrm{cr}}$) for $%
S=1$ in Fig. \ref{fig3} becomes, at large $\Omega $, asymptotically linear
and parallel to the upper existence border, $\mu _{\mathrm{cr}}\approx \mu
_{0}(\Omega )-0.724=1.276+\Omega /2$. On the other hand, the stability
region for the fundamental trapped modes expands at $\Omega \rightarrow
\infty $, because the entire fundamental family is stable in the 2D limit,
see below too.

\begin{figure}[t]
\begin{center}
\includegraphics[width=10cm]{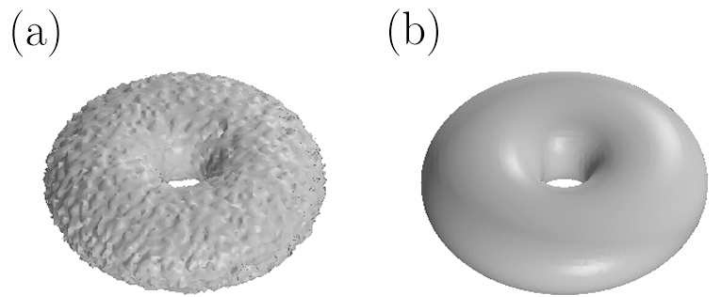}
\end{center}
\caption{Self-cleaning of a stable 3D vortex torus with $S=1$ in the
isotropic model ($\Omega =1$) after the application of a random perturbation
at the amplitude level of $10\%$, as per Ref. \protect\cite{Dum3D}. Panels
(a) and (b) display. severally, the initial shape of the vortex at $t=0$,
and the shape at $t=120$. The unperturbed vortex has chemical potential $%
\protect\mu =2$ and norm $N=12.55$.}
\label{fig4}
\end{figure}

The predictions for the stability based on the computation of the stability
eigenvalues are corroborated by direct simulations of Eq. (\ref{GPE}),
starting with stationary modes to which small arbitrary perturbations are
added. The robustness of stable vortices is illustrated by Fig. \ref{fig4},
which demonstrates that they absorb the perturbations and clean up
themselves.

Lastly, it is relevant to stress that\ stable matter-wave solitons, created
experimentally in condensates of $^{7}$Li \cite{Randy-NJP} and $^{85}$Rb
\cite{Rb85} atoms loaded in cigar-shaped traps, may be well described by the
above solutions for $\Omega \simeq 0.1-0.05$. On the other hand, stable
solitons were also observed in the post-collapse state of the $^{85}$Rb
condensate trapped in the HO potential with \ $\Omega \approx 0.4$, which is
much closer to the isotropic configuration.

\subsection{2D fundamental and vortex trapped modes}

As said above, taking the limit of $\Omega \rightarrow \infty $ in HO
potential (\ref{HO}) leads to the shrinkage of the 3D model into its 2D
simplification, with Eqs. (GPE), (\ref{u}), (\ref{E}), (\ref{N}), (\ref{M}),
and (\ref{growth}) carrying over into their 2D counterparts, and the 2D HO
potential taking the form of $U\left( x,y\right) =(1/2)\left(
x^{2}+y^{2}\right) $. In particular, stationary solutions of the 2D version
of Eq. (\ref{GPE}) are looked for as $\psi =\exp \left( -i\mu t+iS\theta
\right) R(\rho )$, cf. Eq. (\ref{vort}), where $\mu $ is realized as the
chemical potential of the 2D model, i.e., its limit form corresponding to
the linear equation is%
\begin{equation}
\mu _{0}^{\mathrm{(2D)}}=S+1,  \label{S+1}
\end{equation}%
cf. Eq. (\ref{bound}).

\begin{figure}[t]
\includegraphics[width=10cm]{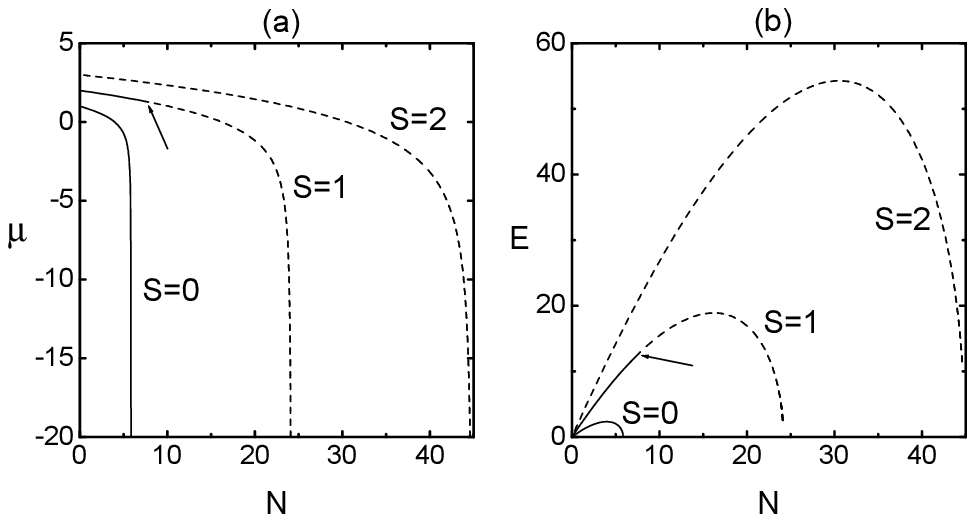}
\caption{The chemical potential $\protect\mu $ (a) and the energy $E$ (b)
versus norm $N$ for the 2D trapped modes with vorticities $S=0,1,$ and $2$,
as per Ref. \protect\cite{Dum2D}. Solid and dashed lines denote stable and
unstable solutions. The arrow indicates the point where the vortices $S=1$
loose their stability.}
\label{fig5}
\end{figure}

The shape and stability of the 2D trapped modes was studied in a number of
works \cite{2D,Dum2D}. The results, produced by the computation of the
stability eigenvalues in the framework of the 2D version of Eq. (\ref{growth}%
), are summarized in Fig. \ref{fig5}, which demonstrates that the family of
the fundamental ($S=0$) trapped modes is \emph{entirely stable} in its
existence region,%
\begin{equation}
0\leq N<N_{\max }^{(S=0)}\approx 5.85,  \label{Nmax}
\end{equation}%
where $N_{\max }^{(S=0)}$ is the numerically found value of the norm of the
Townes soliton, which determines the threshold for the onset of the critical
collapse in the 2D NLSE \cite{collapse} (the variational method makes it
possible to produce an analytical approximation, $N_{\max }^{(S=0)}\approx
2\pi $, with a relative error $\approx 7\%$ \cite{Anderson}). In the free
space (in the absence of any trapping potential), the family of the Townes
solitons is degenerate, in the sense that they all have the single value of
the norm, which is exactly equal to $N_{\max }^{(S=0)}$, independently of
the soliton's chemical potential, $\mu $. The degeneracy is a consequence of
the specific scaling invariance of the NLSE in 2D \cite{collapse}. The
trapping potential introduces a characteristic spatial scale (the standard
HO length, in dimensional units, or merely the spatial period, in the case
of lattice potentials), which breaks the scaling invariance and thus lifts
the degeneracy, making $N$ a function of $\mu $. In fact, Eq. (\ref{Nmax})
demonstrates that the degeneracy is lifted so that the soliton's norm falls
\emph{below} the collapse-onset threshold. This circumstance lends the
trapped modes with $S=0$ protection against the collapse, i.e.,, stability,
and actually makes them the system's GS, which did not exist in the absence
of the trapping potential. This is a major difference from the stabilization
mechanism provided by the trapping potential in 3D, where, as said above,
the GS cannot exist (as the supercritical collapse is possible at any value
of the norm), and only metastability of the trapped modes is possible.

The family of Townes solitons with embedded vorticity $S=1$ is degenerate
too in the free space, being pinned to the single value of the norm, $N_{%
\mathrm{\max }}^{(S=1)}\approx 24.1$ \cite{Minsk}. As seen in Fig. \ref{fig5}%
, the HO trapping potential stabilizes them in interval%
\begin{equation}
0\leq N<7.79\approx 0.32N_{\mathrm{\max }}^{(S=1)},  \label{main}
\end{equation}%
the corresponding stability interval in terms of the chemical potential
being $1.27575\equiv \mu _{\mathrm{cr}}<\mu \equiv \mu _{0}^{\mathrm{(2D)}%
}(S=1)\equiv 2$ (the right edge of the interval is determined by Eq. (\ref%
{S+1})). As for the vortices with $S\geq 2$, they remain completely
unstable, as in the 3D model.

The stability of the trapped vortex modes with $S=1$, predicted through the
computation of the corresponding eigenvalues, was corroborated by direct
simulations of the perturbed evolution, see a typical example in Fig. \ref%
{fig6}.
\begin{figure}[t]
\includegraphics[width=10cm]{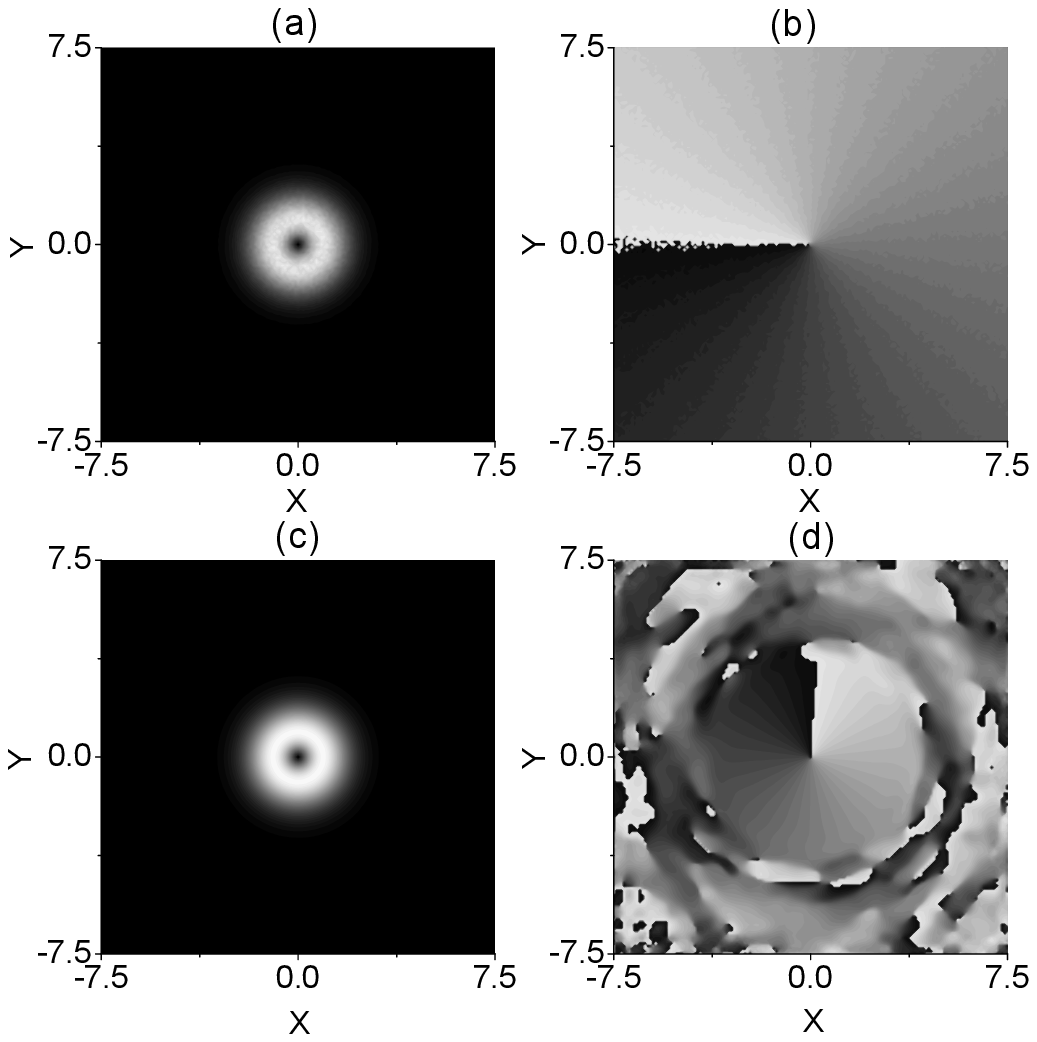}
\caption{Grey-scale plots illustrating recovery of a perturbed 2D trapped
vortex mode with $S=1$ for $\protect\mu =1.4$ and $N=6.64$, which belongs to
stability interval (\protect\ref{main}). (a,b) Intensity and phase fields in
the initial vortex, contaminated by a random-noise perturbation. (c,d) The
same in the self-cleaned vortex mode at $t=120$, as per Ref. \protect\cite%
{Dum2D}.}
\label{fig6}
\end{figure}

\begin{figure}[t]
\includegraphics[width=10cm]{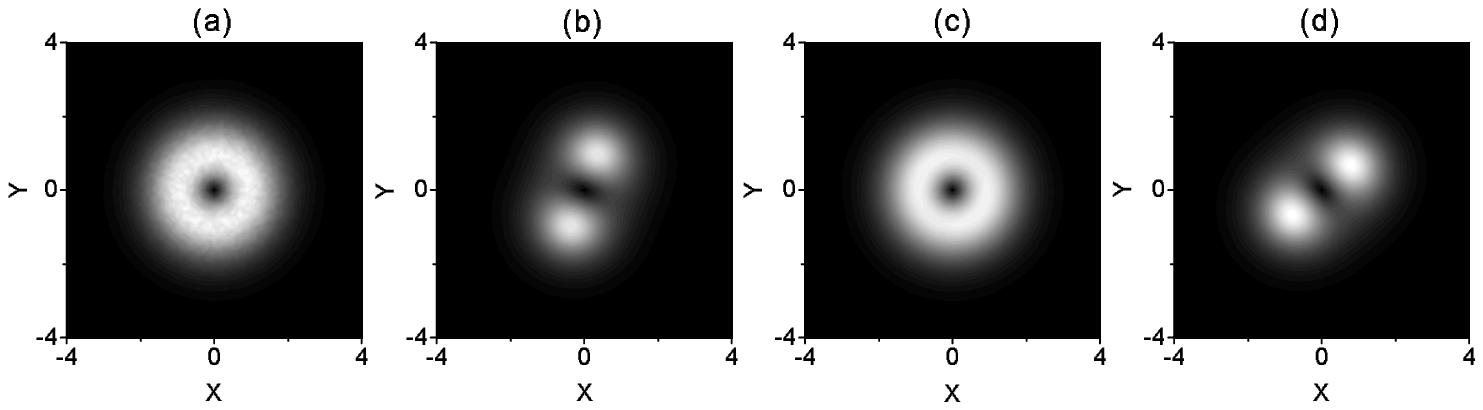}
\caption{Periodic evolution of the unstable vortex with $S=1$, $\protect\mu %
=1.2$ and $N=8.48$, which belongs to interval (\protect\ref{intermediate})
of the semi-stability, as per Ref. \protect\cite{Dum2D} In this dynamical
regime, the trapped mode periodically splits into two fragments and
recombines, while its vorticity is conserved: (a) $t=0$, (b) $t=100$, (c) $%
t=140$, and (d) $t=180$.}
\label{fig7}
\end{figure}

The direct simulations also reveal an interesting dynamical regime of the
trapped vortices with $S=1$ in interval
\begin{equation}
0.32N_{\mathrm{\max }}^{(S=1)}\approx 7.79<N<10.30\approx 0.43N_{\mathrm{%
\max }}^{(S=1)},  \label{intermediate}
\end{equation}%
adjacent to one given by Eq. (\ref{main}). In this interval, the evolution
of the unstable vortex is \emph{regular} (time-periodic), as shown in Fig. %
\ref{fig7}: it spits into two fragments which then recombine back into the
vortex. The splitting-recombination cycles recur periodically, keeping the
vorticity of the configuration. In this dynamical regime, the trapped
vortices may be considered as \emph{semi-stable} modes.

The vortices with still larger values of the norm, $N>10.30$, also split
into two fragments, which, however, fail to recombine. Instead, each one
quickly blows up, i.e., collapses. Note that the corresponding effective
collapse threshold, $N\approx 10.30$, is close to the double collapse
threshold of the fundamental solitons, $2N_{\max }^{(S=0)}\approx 11.70$,
which is consistent with the observation that the two fragments tend to
suffer the intrinsic collapse if they do not recombine.

A noteworthy generalization of the above analysis\ was performed for a
system of two nonlinearily coupled fields, which may be realized as a binary
BEC, or as the copropagation of two optical beams in a bulk waveguide \cite%
{Brtka}. The respective coupled 2D GPE/NLSE system is
\begin{subequations}
\begin{align}
i\frac{\partial \psi _{1}}{\partial t}& =\left[ -\frac{1}{2}\nabla ^{2}+%
\frac{1}{2}\left( x^{2}+y^{2}\right) -\left( |\psi _{1}|^{2}+\eta \left\vert
\psi _{2}\right\vert ^{2}\right) \right] \psi _{1},  \notag \\
&  \label{system} \\
i\frac{\partial \psi _{2}}{\partial t}& =\left[ -\frac{1}{2}\nabla ^{2}+%
\frac{1}{2}\left( x^{2}+y^{2}\right) -\left( |\psi _{2}|^{2}+\eta \left\vert
\psi _{1}\right\vert ^{2}\right) \right] \psi _{2},  \notag
\end{align}%
where $\eta $ is the relative strength of the attraction ($\eta >0$) or
repulsion ($\eta <0$) between the components. This system gives rise to
composite modes with \emph{hidden vorticity}, formed by the components with
equal norms $N/2$ and opposite vorticities, $S_{1,2}=\pm 1$. Such modes are
stable, roughly, in the parameter region with $0\leq N<8,~-1.0\leq \beta
\leq 0.2$.

\section{Stabilization of 2D and 3D semi-vortex and mixed-mode solitons by
the spin-orbit coupling}

\subsection{The models}

In this section, in contrast to the previous one, which outlined well-known
results, established between 10 and 20 years ago \cite{2D,Dum2D,Dum3D}, very
recent findings are summarized, which put forward a completely new approach
to the stabilization of vorticity-carrying 2D and 3D solitons in \emph{free
space}, i.e., without the use of any external potential. The novel
stabilization mechanism can be realized in recently introduced models of
binary (pseudo-spinor) atomic BEC with the SOC.

A great deal of attention has been lately drawn to the use of ultracold
quantum gases, both bosonic and fermionic, as \textit{simulators} of
various\ fundamental effects that were previously predicted and/or
discovered experimentally in much more complex settings of condensed-matter
physics \cite{simulator}. In particular,\ much interest has been recently
attracted to the implementation of the (pseudo-) SOC in atomic BEC, as an
efficient emulation of the fundamental SOC and Zeeman-splitting effects in
semiconductors, where SOC is induced by the\ coupling of the electron's
magnetic moment to the magnetic field generated by the intrinsic
electrostatic field of the crystal in the reference frame moving along with
the electron \cite{Dresselhaus,Rashba}.The experimental implementation of
the pseudo-SOC was proposed \cite{Campbell} and realized in the condensate
of $^{87}$Rb atoms, using appropriately designed laser illumination and
magnetic fields \cite{socbec}. Parallel to that, a large number of
theoretical studies on this topic have been carried out \cite{theory-SOC}-%
\cite{Fukuoka2}, \cite{we,HP}. Reviews of this field are available too \cite%
{rf:1}. The SOC emulation in the atomic condensate is provided by mapping
the spinor wave function of semiconductor electrons onto a two-component
pseudo-spinor wave function of the binary BEC composed of atoms in two
different hyperfine states. Namely, a pair of states of the $^{87}$Rb atom, $%
\left\vert \psi _{+}\right\rangle =\left\vert F=1,m_{F}=0\right\rangle $ and
$\left\vert \psi _{-}\right\rangle =\left\vert F=1,m_{F}=-1\right\rangle $,
were used to map the spin-up and spin-down electron's wave function into
them \cite{socbec}. Thus, the dynamics of the fermionic wave functions of
electrons in the semiconductor can be mapped into the mean-field dynamics of
the bosonic gas.

The consideration of the interplay of the SOC, which is, essentially, a
linear mixing between the two components of the spatially inhomogeneous
binary BEC, and the intrinsic nonlinearity in the bosonic condensate has
made it possible to predict diverse nonlinear patterns strongly affected or
created by the SOC, including a variety of 1D solitons \cite{Konotop}, 2D
gap solitons supported by optical-lattice potentials \cite{gap-sol}, and 2D
vortices and vortex lattices, in forms specific to the SO-coupled BEC \cite%
{Fukuoka}.

Here, we will consider the 2D model of the binary SOC BEC based on the
following system of based on the following system of the coupled GPEs for up
and down components of the spinor wave function,$\Psi \equiv \left\{ \psi
_{+},\psi _{-}\right\} $ \cite{we}:
\end{subequations}
\begin{gather}
\left[ i\frac{\partial }{\partial t}+\frac{1}{2}\nabla ^{2}+i\lambda \left(
-\sigma _{y}\frac{\partial }{\partial x}+\sigma _{x}\frac{\partial }{%
\partial y}+\Omega \sigma _{z}\right) \right.   \notag \\
\left. +\left(
\begin{array}{cc}
|\psi _{+}|^{2}+\eta |\psi _{-}|^{2} & 0 \\
0 & |\psi _{-}|^{2}+\eta |\psi _{+}|^{2}%
\end{array}%
\right) \right] \left(
\begin{array}{c}
\psi _{+} \\
\psi _{-}%
\end{array}%
\right) =0,  \label{R2D}
\end{gather}%
where $\nabla ^{2}=\partial ^{2}/\partial x^{2}+\partial ^{2}/\partial y^{2}$%
, $\sigma _{x,y,z}$ are the Pauli matrices, $\lambda $ is a real coefficient
of the SOC of the Rashba type (the Dresselhaus coupling, with combination $%
\sigma _{x}\partial /\partial x-\sigma _{y}\partial /\partial y$, instead of
$-\sigma _{y}\partial /\partial x+\sigma _{x}\partial /\partial y$ in Eq. (%
\ref{R2D}), is not considered here, as it tends to destroy 2D solitons,
rather than to create them \cite{Sherman}), $\eta $ is the relative strength
of the cross-attraction between the components (cf. Eq. (\ref{system})),
under the condition that the strength of the self-attraction is normalized
to be $1$, and $\Omega $ is the strength of the Zeeman splitting.
Coefficient $1/\lambda $ has the dimension of length, defining a fixed scale
which breaks the above-mentioned scale invariance of the NLSE/GPE in the
free 2D space. This effect makes it possible to create stable solitons with
norms falling below the collapse threshold, see below.

Stationary solutions of Eq. (\ref{R2D}) for 2D solitons with real chemical
potential $\mu $ are looked for as $\psi _{\pm }=\exp \left( -i\mu t\right)
u_{\pm }\left( x,y\right) $, where complex stationary wave functions are
determined by equations%
\begin{gather}
\mu u_{+}=-\frac{1}{2}\nabla ^{2}u_{+}-(|u_{+}|^{2}+\gamma |u_{-}|^{2})u_{+}+
\notag \\
\left( \frac{\partial u_{-}}{\partial x}-i\frac{\partial u_{-}}{\partial y}%
\right) -\Omega u_{+},  \label{+} \\
\mu u_{-}=-\frac{1}{2}\nabla ^{2}u_{-}-(|u_{-}|^{2}+\gamma |u_{+}|^{2})u_{-}-
\notag \\
\left( \frac{\partial u_{+}}{\partial x}+i\frac{\partial u_{+}}{\partial y}%
\right) +\Omega u_{-}.  \label{-}
\end{gather}%
Dynamical invariants of Eqs. (\ref{R2D}) are the total norm, energy, and
linear momentum:%
\begin{equation}
N=\int \int \left( \left\vert \psi _{+}\right\vert ^{2}+\left\vert \psi
_{-}\right\vert ^{2}\right) dxdy\equiv N_{+}+N_{-},  \label{norm2D}
\end{equation}%
\begin{gather}
E=\int \int \left\{ \frac{1}{2}\left( |\nabla \psi _{+}|^{2}+|\nabla \psi
_{-}|^{2}\right) -\frac{1}{2}\left( |\psi _{+}|^{4}+|\psi _{-}|^{4}\right)
-\gamma |\psi _{+}|^{2}|\psi _{-}|^{2}+\Omega \left( \left\vert \psi
_{-}\right\vert ^{2}-\left\vert \psi _{+}\right\vert ^{2}\right) \right.
\notag \\
\left. +\lambda \left[ \psi _{+}^{\ast }\left( \frac{\partial \psi _{-}}{%
\partial x}-i\frac{\partial \psi _{-}}{\partial y}\right) +\psi _{-}^{\ast
}\left( -\frac{\partial \psi _{+}}{\partial x}-i\frac{\partial \psi _{+}}{%
\partial y}\right) \right] \right\} dxdy,  \label{ER}
\end{gather}%
\begin{equation}
\mathbf{P}=i\int \int \left( \psi _{+}^{\ast }\nabla \psi _{+}+\psi
_{-}^{\ast }\nabla \psi _{-}\right) dxdy.  \label{P}
\end{equation}%
It is relevant to note that a solution to the linearized version of Eq. (\ref%
{R2D}), in the form of $\psi _{\pm }\left( \mathbf{r},t\right) =\psi _{\pm
}^{(0)}\exp \left( i\mathbf{k}\cdot \mathbf{r}-i\omega t\right) $, gives
rise to two branches of the spectrum,%
\begin{equation}
\omega =\left\{ \frac{k^{2}}{2}+\sqrt{\lambda ^{2}k^{2}+\Omega ^{2}},\frac{%
k^{2}}{2}-\sqrt{\lambda ^{2}k^{2}+\Omega ^{2}}\right\} ,
\end{equation}%
which is gapless in the case of $\Omega =0$, and features a gap, $0\leq
\omega \leq \Omega $, if the Zeeman splitting is included.

The consistent derivation of the effective 2D SOC\ model from the full 3D
system of GPEs may give rise to the 2D equations with nonpolynomial
nonlinearity, as a generalization of the cubic terms in Eq. (\ref{R2D}) \cite%
{Wesley}. This generalized system also creates stable solitons in the 2D
free space. Another relevant generalization addresses a model of a dual-core
nonlinear coupler in optics, where the SOC is emulated by temporal
dispersion of the linear inter-core coupling \cite{Kart}. It was
demonstrated that this model gives rise to stable 2D spatiotemporal solitons
(``light bullets").

The 3D model of the SOC system will be taken as per Ref. \cite{HP}:%
\begin{gather}
\left[ i\frac{\partial }{\partial t}+\frac{1}{2}\nabla ^{2}+i\lambda \nabla
\cdot {\boldsymbol{\sigma }}\right.  \notag \\
\left. +\left(
\begin{array}{cc}
|\psi _{+}|^{2}+\eta |\psi _{-}|^{2} & 0 \\
0 & |\psi _{-}|^{2}+\eta |\psi _{+}|^{2}%
\end{array}%
\right) \right] \left(
\begin{array}{c}
\psi _{+} \\
\psi _{-}%
\end{array}%
\right) =0,  \label{3D}
\end{gather}%
where $\lambda $ is again the SOC coefficient, and the 3D matrix vector is ${%
\boldsymbol{\sigma =}}\left\{ \sigma _{x},\sigma _{y},\sigma _{z}\right\} $.
The 3D system conserves the norm and momentum, given by 3D counterparts of
Eqs. (\ref{norm2D}) and (\ref{P}), and energy%
\begin{gather}
E_{\mathrm{tot}}=E_{\mathrm{kin}}+E_{\mathrm{int}}+E_{\mathrm{SOC}}\,,
\label{eq1} \\
E_{\mathrm{kin}}=\frac{1}{2}\int \int \int \,\left( |\nabla \psi
_{+}|^{2}+|\nabla \psi _{-}|^{2}\right) dxdydz,~  \notag \\
E_{\mathrm{int}}=-\frac{1}{2}\int \int \int \,\left( |\psi _{+}|^{4}+|\psi
_{-}|^{4}+2\eta |\psi _{+}\psi _{-}|^{2}\right) dxdydz\,  \notag \\
E_{\mathrm{SOC}}=-i\lambda \int \int \int \,\Psi ^{\dag }\left( \nabla \cdot
{\boldsymbol{\sigma }}\right) \Psi dxdydz.  \notag
\end{gather}%
The Zeeman splitting is not included into the 3D system, but it may be added
to it.

\subsection{Stable 2D solitons: standing and mobile semi-vortices and mixed
modes}

\subsubsection{2D semi-vortices (at $\Omega =0$)}

First, we note that Eq. (\ref{R2D}) with $\Omega =0$ (in the absence of the
Zeeman splitting) admits stationary solutions with chemical potential $\mu $%
, written in terms of the polar coordinates:%
\begin{equation}
\psi _{+}\left( x,y,t\right) =e^{-i\mu t}f_{1}(\rho ),~\psi _{-}\left(
x,y,t\right) =e^{-i\mu t+i\theta }\rho f_{2}(\rho ),  \label{frf}
\end{equation}%
where real functions $f_{1,2}\left( \rho \right) $ take finite values and
have zero derivatives at $\rho =0$, and feature the following asymptotic
form at $\rho \rightarrow \infty $:%
\begin{equation}
f_{1}\approx F\rho ^{-1/2}e^{-\sqrt{-2\mu -\lambda ^{2}}\rho }\cos \left(
\lambda r+\delta \right) ,~f_{2}\approx -Fr^{-3/2}e^{-\sqrt{-2\mu -\lambda
^{2}}\rho }\sin \left( \lambda r+\delta \right) ,  \label{asympt}
\end{equation}%
with constants $F$ and $\delta $. As it follows from Eq. (\ref{asympt}), the
solutions are exponentially localized, as solitons, at
\begin{equation}
\mu <-\lambda ^{2}/2.  \label{mulambda}
\end{equation}

Solutions (\ref{frf})\ are built as bound states of a fundamental
(zero-vorticity, $S_{+}=0$) soliton in component $\psi _{+}$ and a solitary
vortex, with vorticity $S_{-}=1$, in $\psi _{-}$, therefore composite modes
of this type are called \textit{semi-vortices} (SVs) \cite{we}. The
invariance of Eq. (\ref{GPE}) with respect to transformation
\begin{equation}
\psi _{{\pm }}\left( r,\theta \right) \rightarrow \psi _{{\mp }%
}\left( r,\pi -\theta \right)   \label{transform}
\end{equation}%
gives rise to a semi-vortex which is a mirror image of (\ref{frf}), with $%
\left( S_{+}=0,S_{-}=1\right) $ replaced by $\left( S_{+}=-1,S_{-}=0\right) $%
:%
\begin{equation}
\psi _{+}\left( x,y,t\right) =-e^{-i\mu t-i\theta }\rho f_{2}(\rho ),~\psi
_{-}=e^{-i\mu t}f_{1}(\rho ).  \label{mirror}
\end{equation}

\begin{figure}[b]
\begin{center}
\includegraphics[height=3.5cm]{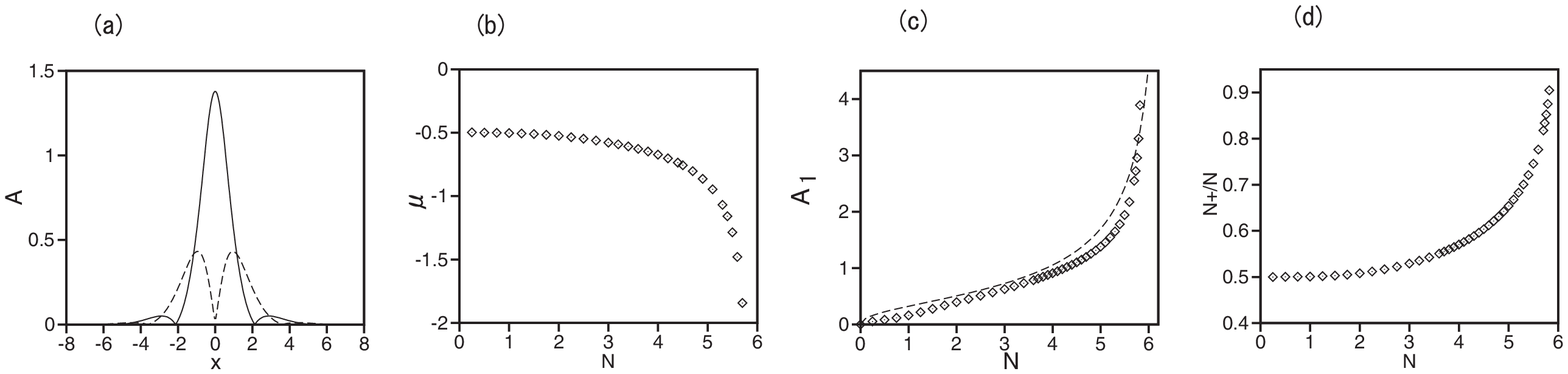}
\end{center}
\caption{(a) Cross sections of the fundamental, $\left\vert \protect\psi %
_{+}\left( x,y\right) \right\vert $, and vortical, $\left\vert \protect\psi %
_{-}\left( x,y\right) \right\vert $, components of a 2D semi-vortex with $%
N=5 $, along axis $y=0$, are shown by continuous and dashed lines,
respectively. (b) Chemical potential $\protect\mu $ vs. norm $N$ for the
family of semi-vortices. (c) Comparison of the numerically found amplitude
of the semi-vortex solution, $|\protect\phi _{+}(x=0,y=0)|$ (the chain of
rhombuses), and its counterpart $A_{1}$, as predicted by the variational
approximation based on ansatz (\protect\ref{frf}) (the dashed curve). (d)
Ratio $N_{+}/N$ as a function of $N$, for the semi-vortex family. In this
figure, $\protect\lambda =1$ and $\protect\eta =\Omega =0$ are fixed in Eq. (%
\protect\ref{R2D}).}
\label{fig8}
\end{figure}

Numerically, stable SVs were generated, as solutions to Eq. (\ref{GPE}) by
means of imaginary-time simulations \cite{im-time}, starting from the
Gaussian input,
\begin{equation}
\psi _{+}^{(0)}=A_{1}\exp \left( -\alpha _{1}\rho ^{2}\right) ,\;\psi
_{-}^{(0)}=A_{2}\rho \exp \left( i\theta -\alpha _{2}\rho ^{2}\right) ,
\label{00}
\end{equation}%
where $A_{1,2}$ and $\alpha _{1,2}>0$ are real constants. A typical example
of the SV is displayed in Fig. \ref{fig8}(a).

Further, Fig. \ref{fig8}(b) represents the family of the SVs, showing
their chemical potential as a function of the norm, cf. Fig. \ref{fig5}(a).
Note that the $\mu (N)$ dependence satisfies the VK criterion, $d\mu /dN<0$,
which is the above-mentioned necessary condition for the stability of
solitary modes supported by the self-attractive nonlinearity. The family of
the SV solitons exists precisely in the interval of norms (\ref{Nmax}),
which, as said above, should secure their stability against the critical
collapse. It is also worthy to note that there is no finite minimum
(threshold) value of $N$ necessary for the existence of the SVs in the free
space. In the limit of $\mu \rightarrow -\infty $, the vortex component of
the SV vanishes, while the fundamental one degenerates into the usual Townes
soliton, with $N=N_{\max }^{(S=0)}$, as shown by means of the dependence of
ratio $N_{+}/N$ on $N$ in Fig. \ref{fig8}(d).

Systematic real-time simulations confirm the stability of the whole SV
family at $\eta \leq 1$, while they are unstable at $\eta >1$, where,
however, there is another family of stable solitons in the form of \textit{%
mixed modes} (MMs), see below. In fact, the SVs at $\eta \leq 1$ and MMs at $%
\eta \geq 1$ are the first ever found examples of stable solitons supported
by the cubic self-attractive nonlinearity in the free 2D space.

The wave form (\ref{00}) was used not only as the input for the
imaginary-time simulations, but also as an ansatz for the application of the
VA. The substitution of the ansatz into expression (\ref{ER}) for the energy
yields
\begin{equation}
E_{\mathrm{SV}}=\pi \left[ \frac{A_{1}^{2}}{2}-\frac{A_{1}^{4}}{8\alpha _{1}}%
+\frac{A_{2}^{2}}{2\alpha _{2}}-\frac{A_{2}^{4}}{64\alpha _{2}^{3}}-\frac{%
\gamma A_{1}^{2}A_{2}^{2}}{4(\alpha _{1}+\alpha _{2})^{2}}+\frac{4\lambda
A_{1}A_{2}\alpha _{1}}{(\alpha _{1}+\alpha _{2})^{2}}\right] ,  \label{Eans}
\end{equation}%
while norm (\ref{norm2D}) of the ansatz is $N=\pi \left[ A_{1}^{2}/\left(
2\alpha _{1}\right) +A_{2}^{2}/\left( 4\alpha _{2}^{2}\right) \right] .$
Then, values of amplitudes $A_{1},A_{2}$ and inverse squared widths $\alpha
_{1}$,$\alpha _{2}$ of the ansatz are predicted by the minimization of $E$
with respect to the variational parameters, $\partial E_{\mathrm{semi}%
}/\partial \left( A_{1,2},\alpha _{1,2}\right) =0$, under the constraint
that $N$ is kept constant. Figure \ref{fig8}(c) displays the comparison
of the so predicted amplitude $A_{1}$ and the maximum value of $|\psi
_{+}\left( x,y\right) |$ obtained from the numerical solution at $\eta =0$.
It is seen that the accuracy of the VA is quite reasonable.

Very recently, the study of 2D\ SV solitons was extended to a system with
long-range anisotropic cubic interactions, mediated by dipole-dipole forces
in a bosonic gas composed of atoms carrying magnetic moments \cite{Raymond}.
An essentially novel feature found in the nonlocal model is a new parameter
of the family of SV solitons, \textit{viz}., a spontaneous shift of the
pivot of the vortical component ($\psi _{-}$) with respect to its
zero-vorticity counterpart ($\psi _{+}$). Those solitons also demonstrate an
essentially different mobility scenario from the one outlined below for the
present model: they respond to an applied kick by drift in the opposite
direction (with an effective negative mass) along a spiral trajectory.

\subsubsection{2D mixed modes (at $\Omega =0$)}

Aside from the SVs, the same SOC system (\ref{R2D}) gives rise to another
type of vorticity-carrying solitons, in the form of MMs, which combine terms
with zero and unitary vorticities, $\left( S=0,S=-1\right) $ and $\left(
S=0,S=+1\right) $, in the spin-up and spin-down components, $\psi _{+}$ and $%
\psi _{-}$. Numerically, the MM can be produced by imaginary-time
simulations initiated by the following input, which may also serve as the
variational ansatz:
\begin{eqnarray}
\psi _{+}^{(0)} &=&A_{1}\exp \left( -\alpha _{1}\rho ^{2}\right) -A_{2}\rho
\exp \left( -i\theta -\alpha _{2}\rho ^{2}\right) ,  \notag \\
\psi _{-}^{(0)} &=&A_{1}\exp \left( -\alpha _{1}\rho ^{2}\right) +A_{2}\rho
\exp \left( i\theta -\alpha _{2}r^{2}\right) .  \label{mixed}
\end{eqnarray}%
In fact, the MM may be considered as a superposition of the SV\ (\ref{frf})
and its mirror image (\ref{mirror}). Accordingly, symmetry reflection (\ref%
{transform}) transforms the MM into itself.

\begin{figure}[tbp]
\begin{center}
\includegraphics[height=3.5cm]{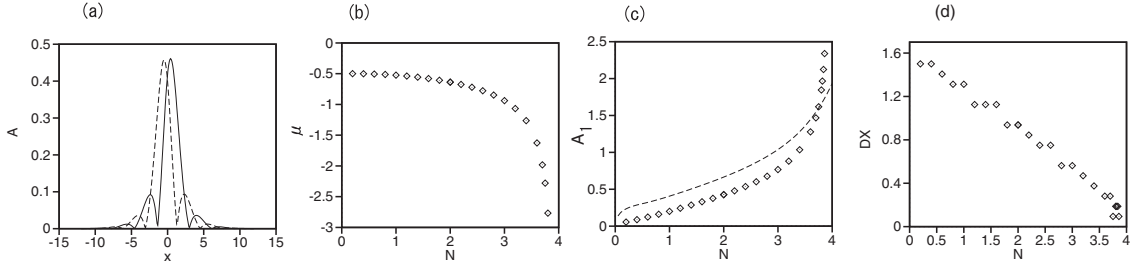}
\end{center}
\caption{Panels (a), (b), and (c) display the same as in Fig. \protect\ref%
{fig8}, but for a 2D mixed-mode soliton at $\protect\eta =2,$ $\protect%
\lambda =1$, and $\Omega =0$ in Eq. (\protect\ref{R2D}). The norm of the
soliton shown in panel (a) is $N=5$. The variational approximation,
represented in panel (c), is based on ansatz (\protect\ref{mixed}). (d)
Separation $\mathrm{DX}$ between peak positions of components $|\protect\psi %
_{+}|$ and $|\protect\psi _{-}|$ vs $N$. }
\label{fig9}
\end{figure}

A typical example of the MM state is displayed in Fig.~\ref{fig9}(a). Note
that peak positions of the two components, $|\psi _{+}\left( x,y\right) |$
and $|\psi _{-}\left( x,y\right) |$, in this state are separated along $x$,
Fig. \ref{fig9}(d) showing the separation ($\mathrm{DX}$) as a function of
the norm. For a small amplitude of the vortex component, $A_{2}$, Eq. (\ref%
{mixed}) yields $\mathrm{DX}\approx A_{2}/\left( \alpha _{1}A_{1}\right) $.

The $\mu (N)$ dependence for the MM family is displayed in Fig. \ref{fig9}%
(b), which shows that the VK criterion holds in this case too, and, as well
as SVs, the MMs do not have any threshold value of $N$ necessary for their
existence. The family exists in the interval of $N<\tilde{N}_{\max
}^{(S=0)}(\eta )=2N_{\max }^{(S=0)}/(1+\eta )$, where $N_{\max }^{(S=0)}$ is
the critical norm (\ref{Nmax})\ corresponding to the Townes solitons. In the
limit of $N\rightarrow \tilde{N}_{\max }^{(S=0)}(\eta )$ the vortex
components vanish in the MM, and it degenerates into a two-component Townes
soliton, cf. the degeneration of the SV in the limit of $N\rightarrow
N_{\max }^{(S=0)}$, which was shown in Figs. \ref{fig8}(b,d). The separation
between peaks of the two components vanishes in this limit too, see Fig. \ref%
{fig9}(d).

The VA for the MM solutions is generated by the insertion of expression (\ref%
{mixed}), as the variational ansatz, into energy functional (\ref{ER}),
which yields
\begin{equation}
E_{\mathrm{MM}}=\pi \left[ A_{1}^{2}+\frac{A_{2}^{2}}{\alpha _{2}}-(1+\gamma
)\left( \frac{A_{1}^{4}}{4\alpha _{1}}+\frac{A_{2}^{4}}{32\alpha _{2}^{3}}%
\right) -\frac{A_{1}^{2}A_{2}^{2}}{(\alpha _{1}+\alpha _{2})^{2}}+\frac{%
8\lambda A_{1}A_{2}\alpha _{1}}{(\alpha _{1}+\alpha _{2})^{2}}\right] ,
\label{Emix}
\end{equation}%
the total norm of the ansatz being $N=\pi \left[ A_{1}^{2}/\alpha
_{1}+A_{2}^{2}/\left( 2\alpha _{2}^{2}\right) \right] $. Numerical solution
of the respective variational equations, which minimize energy (\ref{Emix})
under the constraint of the fixed norm, produces results which are compared
in Fig. \ref{fig9}(c) with their numerically found counterparts. The VA for
the MM solution is less accurate than for its VA counterpart, cf. Fig. \ref%
{fig8}(c), but it is usable too.

Direct simulations demonstrate that the MMs are unstable at $\eta <1$, and
stable at $\eta \geq 1$, i.e., precisely in the regions where the SVs are,
severally, stable and unstable (exactly at $\eta =1$, both the SV and MM
solutions are stable \cite{Fukuoka2}). This stability switch between the SV
and MM is explained by the comparison of energy (\ref{ER}) for them at equal
values of the norm: the energy is smaller for the SV at $\eta <1$, and for
the MM at $\eta >1$ \cite{we} (similar to what is shown below in Fig. \ref%
{fig13}(a) for the 3D system). Accordingly, the SV and MM realize the
system's stable GS, respectively, at $\eta <1$ and $\eta >1$, while in the
opposite case each soliton species represents an excited state, instead of
the GS, being therefore unstable.

\subsubsection{Effects of the Zeeman splitting}

The above considerations did not include the Zeeman splitting, $\Omega \neq
0 $. In the presence of moderately strong splitting, with $0<\Omega <\lambda
^{2}$, the consideration of Eq. (\ref{R2D}), similar to that which produced
the asymptotic form (\ref{asympt}) of the SV soliton at $\Omega =0$,
demonstrates that the SVs exist at%
\begin{equation}
\mu <-(1/2)\left( \lambda ^{2}+\Omega ^{2}/\lambda ^{2}\right) \equiv \mu
_{\max },  \label{mu<}
\end{equation}%
cf. existence region (\ref{mulambda}) at $\Omega =0$. The asymptotic form of
the SV solution, affected by $\Omega $, is more complex than one given by
Eq. (\ref{asympt}); namely, the radial decay rate and wavenumber in Eq. (\ref%
{asympt}) are replaced as follows:%
\begin{equation}
\sqrt{-2\mu -\lambda ^{2}}\rightarrow \sqrt{-\mu -\lambda ^{2}+\sqrt{\mu
^{2}-\Omega ^{2}}},~\lambda \rightarrow \frac{\sqrt{-2\mu \lambda
^{2}-\lambda ^{4}-\Omega ^{2}}}{\sqrt{-\mu -\lambda ^{2}+\sqrt{\mu
^{2}-\Omega ^{2}}}}.  \label{-->}
\end{equation}

Strong Zeeman splitting, with $\Omega >\lambda ^{2}$, replaces existence
condition (\ref{mu<}) by $\mu <-\lambda ^{2}$. In this case, the SV solitons
keep the asymptotic form (\ref{-->}) in the semi-infinite interval (\ref{mu<}%
) of the chemical potentials, while, in the adjacent finite interval
appearing in this case,
\begin{equation}
\mu _{\max }<\mu <-\lambda ^{2},
\end{equation}%
the SV soliton exhibits a more dramatic change of its asymptotic shape: the
radial wavenumber vanishes, the decay of the solution at $\rho \rightarrow
\infty $ being provided by the following exponential factor,%
\begin{equation}
\exp \left[ -\sqrt{2\left( -\mu -\lambda ^{2}-\sqrt{2\mu \lambda
^{2}+\lambda ^{4}+\Omega ^{2}}\right) }\rho \right]  \label{factor}
\end{equation}%
(at $\mu =\mu _{\max }$, see Eq. (\ref{mu<}), expression (\ref{factor})
matches Eq. (\ref{-->})).

As concerns the switch of the stability between the SV and MM at $\eta =1$,
which was demonstrated above for $\Omega =0$, it is pushed by $\Omega >0$ to
larger values of the relative strength of the cross-attraction, $\eta >1$
\cite{Sherman}. An explicit result can be obtained in the limit of large $%
\Omega $, when Eq. (\ref{+}) demonstrates that the chemical potential is
close to $-\Omega $:
\begin{equation}
\mu =-\Omega +\delta \mu ,~\left\vert \delta \mu \right\vert \ll \Omega .
\label{mu}
\end{equation}%
The spin-down component, $\psi _{-}$, is vanishingly small in this limit,
hence stationary equation (\ref{-}) simplifies to%
\begin{equation}
u_{-}\approx \frac{1}{2\Omega }\left( \lambda \frac{\partial u_{+}}{\partial
x}+i\lambda \frac{\partial u_{+}}{\partial y}\right) ,  \label{u-}
\end{equation}%
where $\left( \Omega -\mu \right) $ is approximately replaced by $2\Omega $,
pursuant to Eq. (\ref{mu}). Then, the substitution of approximation (\ref{u-}%
) into Eq. (\ref{+}) leads to the following equation for $u_{+}$:%
\begin{equation}
\left( \delta \mu \right) u_{+}=-\frac{1}{2}\left( 1-\frac{1}{\Omega }%
\right) \nabla ^{2}u_{+}-|u_{+}|^{2}u_{+}~.  \label{deltamu}
\end{equation}%
By itself, Eq. (\ref{deltamu}) is tantamount to the stationary version of
the 2D NLSE, which gives rise to the Townes solitons in $u_{+}$, while Eq. (%
\ref{u-}) generates a small vortex component of the SV complex in $u_{-}$.
Due to the presence of factor $\left( 1-1/\Omega \right) $ in Eq. (\ref%
{deltamu}) and the smallness of $1/\Omega $, the norm (\ref{norm2D}) of the
corresponding SV complex is%
\begin{equation}
N=\left( 1-\frac{1}{\Omega }\right) N_{\max }^{(S=0)}+\mathcal{O}\left(
\frac{1}{\Omega ^{2}}\right) ,  \label{<}
\end{equation}%
where $N_{\max }^{(S=0)}$ is the standard norm of the Townes soliton (\ref%
{Nmax}), the last term being a contribution from the small vortex component
given by Eq. (\ref{u-}). Equation (\ref{<}) demonstrates that the total norm
of the SV soliton is (slightly) \emph{smaller} than the collapse threshold, $%
N_{\max }^{(S=0)}$, hence the SV soliton is \emph{protected} against the
collapse, still realizing the stable GS of the system.

The lowest-order approximation presented here does not give rise to terms
including the cross-attraction coefficient, $\eta $ from Eq. (\ref{R2D}),
which implies that, in the limit of large $\Omega $, all solitons belong to
the SV type, irrespective of the value of $\eta $, in accordance with the
above-mentioned fact that the increase of $\Omega $ leads to conversion of
the MMs into SVs \cite{Sherman}.

\subsubsection{Mobility of stable 2D solitons}

Although the underlying system (\ref{R2D}) conserves the momentum (\ref{P}),
the SOC terms break the Galilean invariance of the original NLSEs. For this
reason, generating moving solitons from quiescent ones, which were
considered above, is a nontrivial problem. As shown in Ref. \cite{we}, the
system gives rise to the mobility of 2D solitons along the $y$ axis, but not
along $x$. The respective solutions moving at velocity $v_{y}$ can be looked
for as
\begin{equation}
\psi _{\pm }=\exp \left( iv_{y}y-\frac{i}{2}v_{y}^{2}t\right) \phi _{\pm
}(x;y^{\prime }\equiv y-v_{y}t;t)  \label{phi}
\end{equation}%
(Eq. (\ref{phi}) produces the Galilean transform of the wave functions $\psi
_{\pm }$ in Galilean-invariant systems). The substitution of ansatz (\ref%
{phi}) into Eq. (\ref{R2D}) leads to the coupled GPEs in the moving
reference frame, which differ from Eqs. (\ref{R2D}) by the presence of
linear mixing between the two components (here $\nabla ^{2}\equiv \partial
^{2}/\partial x^{2}+\partial ^{2}/\partial \left( y^{\prime }\right) ^{2}$,
and $\Omega =0$ is set):
\begin{eqnarray}
i\frac{\partial \phi _{+}}{\partial t} &=&-\frac{1}{2}\nabla ^{2}\phi
_{+}-(|\phi _{+}|^{2}+\gamma |\phi _{-}|^{2})\phi _{+}^{\prime }+\lambda
\left( \frac{\partial \phi _{-}}{\partial x}-i\frac{\partial \phi _{-}}{%
\partial y^{\prime }}\right) +\lambda v_{y}\phi _{-},  \notag \\
i\frac{\partial \phi _{-}}{\partial t} &=&-\frac{1}{2}\nabla ^{2}\phi
_{-}-(|\phi _{-}|^{2}+\gamma |\phi _{+}|^{2})\phi _{-}-\lambda \left( \frac{%
\partial \phi _{+}}{\partial x}+i\frac{\partial \phi _{+}}{\partial
y^{\prime }}\right) +\lambda v_{y}\phi _{+}.  \label{mix}
\end{eqnarray}%
The same linear mixing can be imposed onto the 2D settings by a GHz wave
coupling the two underlying atomic states \cite{radio}, hence the linear
mixing by itself represents a relevant addition to the model.

Stationary solutions to equations (\ref{mix}), written in the moving
reference frame, can be obtained, as well as in the case of Eqs. (\ref{R2D}%
), by means of the imaginary-time evolution method. In particular, at $\eta
=2$, when the GS is represented by the quiescent MM soliton, its moving
version, which is displayed in Figs. \ref{fig8}(a,b) for $N=3.1$ and $%
v_{y}=0.5$, exists and is stable too. As well as its quiescent counterpart,
this mode features the mirror symmetry between the profiles of $|\phi
_{+}\left( x,y\right) |$ and $|\phi _{-}\left( x,y\right) |$. Figures \ref%
{fig10}(a,b) display a typical example of a stable moving MM, and Fig. \ref%
{fig10}(c) shows the amplitude of the moving soliton, $A=\sqrt{|\phi
_{+}(x=0,y=0)|^{2}+|\phi _{-}(x=0,y=0)|^{2}}$, as a function of $v_{y}$. The
amplitude monotonously decreases with the growth of the velocity, vanishing
at
\begin{equation}
v_{y}=\left( v_{y}\right) _{\max }^{(\mathrm{MM})}\approx 1.8,  \label{Vmax}
\end{equation}%
i.e., the mobile solitons exist in the \emph{limited interval} of the
velocities.
\begin{figure}[tbp]
\begin{center}
\includegraphics[height=3.5cm]{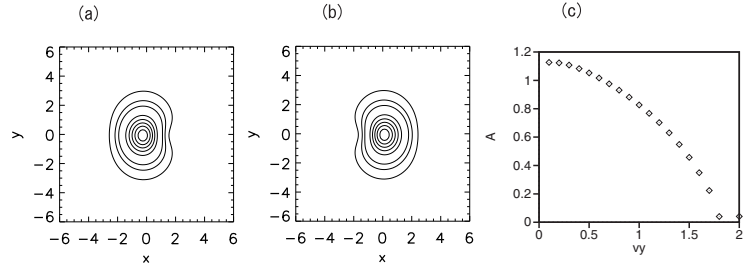}
\end{center}
\caption{Contour plots of $|\protect\psi _{+}\left( x,y\right) |$ (a) and $|%
\protect\psi _{-}\left( x,y\right) |$ (b) of the 2D stable mixed-mode
soliton with norm $N=3.1$, moving at velocity $v_{y}=0.5$ , for $\protect%
\eta =2,\protect\lambda =1,\Omega =0$. (c) The amplitude of the moving
solitons as a function of $v_{y}$.}
\label{fig10}
\end{figure}

The SV solitons may also be made mobile, but in a very narrow interval of
velocities -- e.g., at $v_{y}<\left( v_{y}\right) _{\max }^{(\mathrm{SV}%
)}\approx 0.03$ for $\eta =0,\lambda =1,\Omega =0$, and $N=3.7$. At $%
v_{y}>0.03$, the imaginary-time solution of Eq. (\ref{mix}) with the SV
input converges to stable MM solitons, instead of the SV.

\subsection{3D metastable solitons}

\subsubsection{Analytical considerations}

\begin{figure}[tbp]
\centering\includegraphics[width=0.5\columnwidth]{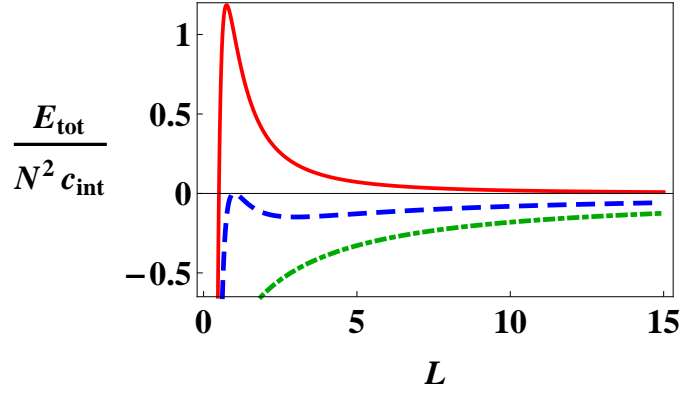}
\caption{(Color online) $E_{\mathrm{tot}}$ as a function of condensate's
size $L$, as per Eq. (\protect\ref{eq2}). The red solid, blue dashed, and
green dot-dashed lines represent the energy's variation when $\protect%
\lambda =0$, and $\protect\lambda >0$ does or does not satisfy condition (%
\protect\ref{eq3}), respectively.}
\label{fig11}
\end{figure}

The creation of metastable 3D solitons in the model based on Eq. (\ref{3D})
can be predicted starting from scaling evaluation of different terms in the
respective energy functional (\ref{eq1}). Assuming that a localized state
has characteristic size $L$ and norm $N$, an estimate for the amplitudes of
the wave function is $A\sim \sqrt{N}L^{-3/2}$. Accordingly, the three terms
in Eq. (\ref{eq1}) scale with $L$ as
\begin{equation}
E_{\mathrm{tot}}/N\sim c_{\mathrm{kin}}L^{-2}-c_{\mathrm{soc}}\lambda
L^{-1}-\left( c_{\mathrm{int}}^{\mathrm{(self)}}+c_{\mathrm{int}}^{\mathrm{%
(cross)}}\eta \right) NL^{-3}\,,  \label{eq2}
\end{equation}%
with positive coefficients $c_{\mathrm{kin}}$, $c_{\mathrm{soc}}$, and $c_{%
\mathrm{int}}^{\mathrm{(self/cross)}}$. As shown in Fig.~\ref{fig11}, Eq.~(%
\ref{eq2}) gives rise to a \emph{local minimum} of $E_{\mathrm{tot}}(L)$ at
finite $L$, provided that
\begin{equation}
0<\lambda N<{c_{\mathrm{kin}}^{2}}/\left[ {3\left( c_{\mathrm{int}}^{\mathrm{%
(self)}}+c_{\mathrm{int}}^{\mathrm{(cross)}}\eta \right) c_{\mathrm{soc}}}%
\right] \,.  \label{eq3}
\end{equation}%
Although this minimum cannot represent the GS, which formally corresponds to
$E_{\mathrm{tot}}\rightarrow -\infty $ at $L\rightarrow 0$ in the collapsed
state, as is suggested by Fig. \ref{fig11} too -- in fact, this means that
the system has no true GS), it corresponds to a self-trapped state which is
stable against small perturbations. Condition (\ref{eq3}) suggests that the
metastable 3D solitons may exist when the SOC term is present, while its
strength $\lambda $ is not too large, $N$ and $\eta $ being not too large
either.

More accurate semi-analytical results can be obtained by means of the VA,
which is based on the following ansatz for the 3D version of the SV mode,
written in cylindrical coordinates $\left( \rho ,z,\theta \right) $, cf. its
2D counterpart (\ref{00}):
\begin{equation}
\left\{ \psi _{+},\psi _{-}\right\} =e^{-i\mu t}\left\{ f_{1}(\rho
,z),e^{i\theta }\,f_{2}(\rho ,z)\right\} \,,  \label{eq5}
\end{equation}%
\begin{equation}
f_{n}=i^{n-1}\left( A_{n}+iB_{n}z\right) \rho ^{n-1}\exp \left( -\alpha
_{n}\rho ^{2}-\beta _{n}z^{2}\right) \;\;(n=1,2)\,,  \label{AB}
\end{equation}%
with real parameters $A_{n},B_{n}$, and $\alpha _{n}>0$, $\beta _{n}>0$. In
particular, amplitudes $A_{n}$ and $B_{n}$ account for possibility of having
even and odd terms in the dependence of the wave functions of the vertical
coordinate, $z$. Further, the ansatz for the 3D version of the MM soliton
may be chosen as a superposition, with mixing angle $\zeta $, of the SV (\ref%
{eq5}) and its flipped counterpart, cf. Eq. (\ref{mirror}) in the 2D
setting:
\begin{equation}
\begin{split}
\psi _{+}& =\left( \cos {\zeta }\right) \,f_{1}(\rho ,z)-\left( \sin {\zeta }%
\right) \,f_{2}^{\ast }(\rho ,z)\,e^{-i\phi }\,, \\
\psi _{-}& =\left( \sin {\zeta }\right) \,f_{1}^{\ast }(\rho ,z)+\left( \cos
{\zeta }\right) \,f_{2}(\rho ,z)\,e^{i\phi }\,.
\end{split}
\label{eq10}
\end{equation}

\begin{figure}[tbh]
\centering\includegraphics[width=0.7\columnwidth]{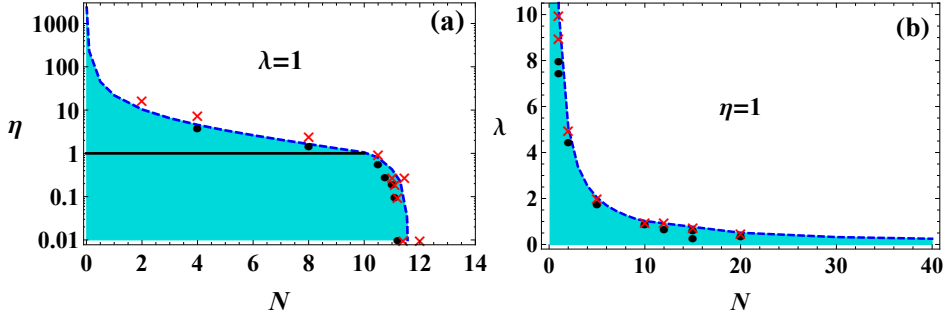}
\caption{(Color online) Stable 3D solitons are predicted by the variational
approximation in blue shaded regions of the parameter planes. In panel (a),
these are semi-vortices at $\protect\eta <1$, and mixed modes at $\protect%
\eta >1$, with the boundary between them depicted by the black solid line.
In (b), the stability area is filled by the solitons of both types, as they
have equal energies at $\protect\eta =1$. The predictions are confirmed by
numerical findings, as indicated by red crosses and black dots, which
indicate, respectively, the absence and presence of stable\ numerically
generated 3D solitons for respective sets of parameters. }
\label{fig12}
\end{figure}

The substitution of \textit{ans\"{a}tze} (\ref{eq5})-(\ref{eq10}) into the
full energy (\ref{eq1})\ and minimizing it with respect to the free
parameters under the constraint of the fixed total norm produces results
summarized in Fig.~\ref{fig12}, in which (meta)stable 3D solitons are
predicted to exist in the shaded areas. It is seen that the solitons indeed
exist, provided that $\lambda $, $N$ and $\eta $ are not too large, in
agreement with the rough prediction of Eq.~(\ref{eq3}). In other words, for
fixed $\lambda $ and $\eta $ the stable 3D\ solitons exist in a finite
interval of the norm,
\begin{equation}
0\leq N\leq N_{\max }\left( \lambda ,\eta \right) \,,  \label{NN}
\end{equation}%
which is qualitatively similar to the situation for the single-component
model stabilized by the HO trapping potential, cf. Figs. \ref{fig1} and \ref%
{fig2}.

\begin{figure}[tbh]
\centering\includegraphics[width=0.65\columnwidth]{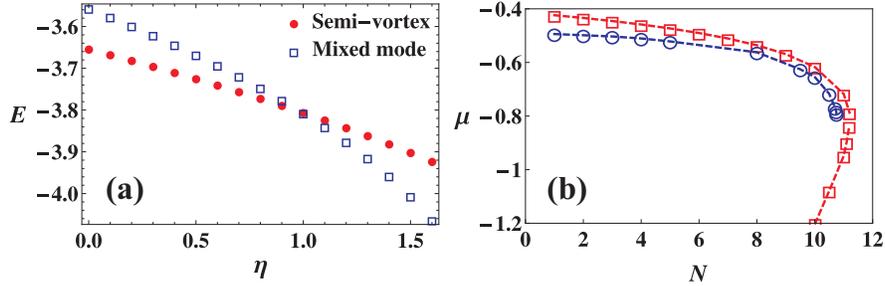}
\caption{(Color online) (a) Energies of the semi-vortices and mixed modes
versus the cross-attraction strenght, $\protect\eta $, as predicted by the
variational approximation for $\protect\lambda =1$ and $N=8$ (as per Ref.
\protect\cite{HP}). The two curves cross at $\protect\eta =1$, where the
energies are equal. (b) Numerically (blue circles) and variationally (red
squares) found chemical potentials versus the norm for the semi-vortices at $%
\protect\lambda =1$ and $\protect\eta =0.3$.}
\label{fig13}
\end{figure}

As shown in Fig.~\ref{fig13}(a), for $\eta <1$ the VA-predicted energy of
the SV is lower than that for the MM, and vice versa for $\eta >1$, similar
to what was found in the 2D model (\ref{R2D}) with $\Omega =0$. Furthermore,
red squares in Fig.~\ref{fig13}(b) depict the VA prediction for the
soliton's chemical potential, $\mu $, plotted as a function of norm $N$ for $%
\lambda =1$ and $\eta =0.3$. These curves clearly demonstrate that, also
similar to what was found in the 2D model (cf. Figs. \ref{fig8}(b) and \ref%
{fig9}(b)), there is no threshold (minimum norm) necessary for the existence
of the stable 3D solitons, which exist up to a $N=N_{\max }$, see Eq. (\ref%
{NN}).

While the stability of the upper branch is consistent with the VK criterion,
$d\mu /dN<0$, the lower branch in Fig. \ref{fig13}(b), which does not
satisfy the VK criterion, represents solitons corresponding to the energy
maximum (instead of the minimum) of the blue dashed curve in Fig.~\ref{fig11}%
, therefore the lower-branch solitons are definitely unstable. In the limit
of $\mu \rightarrow -\infty $, they carry over into the known unstable 3D
solitons of the single GPE \cite{Silberberg}.

\subsubsection{Numerical results}

In the numerical form, stationary 3D solitons were produced by means of
imaginary-time simulations of Eq. (\ref{3D}). Typical examples of the
density profiles in stable SV and MM solitons are displayed in Fig.~\ref%
{fig14}. Naturally, the MM states exhibit a more sophisticated profile.

\begin{figure}[tbh]
\centering\includegraphics[width=0.50\columnwidth]{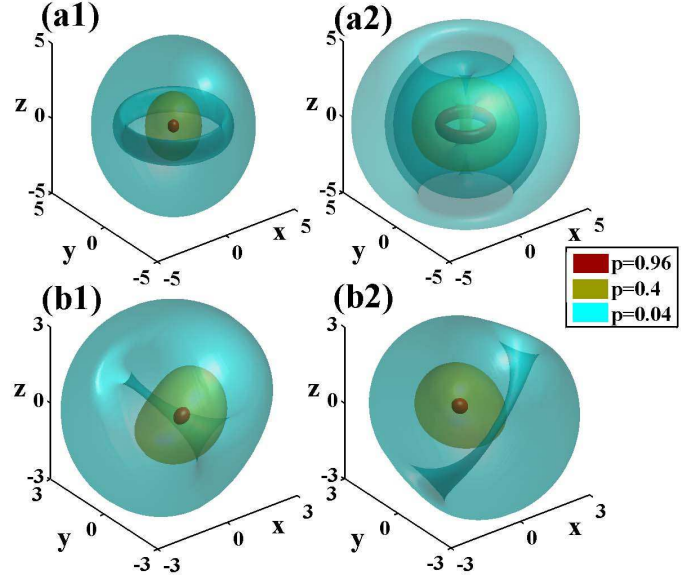}
\caption{(Color online) Density profiles of 3D solitons for $N=8$ and $%
\protect\lambda =1$, as per Ref. \protect\cite{HP}. (a) A semi-vortex for $%
\protect\eta =0.3$, whose fundamental and vortical components, $|\protect%
\psi _{+}|$ and $|\protect\psi _{-}|$, are plotted in (a1) and (a2),
respectively. (b) A mixed mode for $\protect\eta =1.5$, with (b1), (b2)
displaying $|\protect\psi _{+}|$ and $|\protect\psi _{-}|$, respectively. In
each subplot, different colors represent constant-magnitude surfaces, $|%
\protect\psi _{\pm }\left( x,y,z\right) |=\left( 0.96,0.4,0.04\right) \times
|\protect\psi _{\pm }|_{\mathrm{max}}$.}
\label{fig14}
\end{figure}

Symbols in Fig.~\ref{fig12}, which indicate the absence and presence of the
numerically generated stable solitons, are in good agreement with the
predictions of the VA. Further, blue circles in Fig.~\ref{fig13}(b)
represent the numerically obtained soliton's chemical potential, which is
also close to the prediction of the VA. However, the unstable (lower) branch
in Fig. \ref{fig13}(b), predicted by the VA, cannot be produced by the
imaginary-time integration. The stability of the solitons belonging to the
upper branch in Fig.~\ref{fig13}(b) was verified by real-time simulations of
their perturbed evolution. Further, similar to what is shown above for the
2D system in Fig. \ref{fig11}, it was found that stable 3D solitons may be
set in motion in a limited interval of velocities, cf. Eq. (\ref{Vmax}).

Lastly, it was mentioned in Section II that all the higher-order vortex
modes, with $S\geq 2$, are completely unstable in the model with the HO
trapping potential (see Figs. \ref{fig3} and \ref{fig5}). Higher-vorticity
states of both the SV and MM types can be found too as stationary solutions
of Eqs. (\ref{R2D}) \cite{we} and \ref{3D} \cite{HP}, starting from \textit{%
ans\"{a}tze}%
\begin{equation}
\psi _{+}^{(0)}=A_{1}\rho ^{s}\exp \left( is\theta -\alpha _{1}\rho
^{2}\right) ,\;\psi _{-}^{(0)}=A_{2}\rho ^{s+1}\exp \left( i(s+1)\theta
-\alpha _{2}\rho ^{2}\right) ,
\end{equation}%
with $s=1,2,3,...$, for the higher-order 2D SV solitons (cf. Eq. (\ref{00}%
)), and%
\begin{equation}
\left\{ \psi _{+},\psi _{-}\right\} =e^{-i\mu t+is\theta }\left\{ f_{1}(\rho
,z),e^{i\theta }\,f_{2}(\rho ,z)\right\} \,,
\end{equation}%
\begin{equation}
f_{n}=i^{n-1}\left( A_{n}+iB_{n}z\right) \rho ^{s+n-1}\exp \left( -\alpha
_{n}\rho ^{2}-\beta _{n}z^{2}\right) \;\;(n=1,2)\,,
\end{equation}%
for their 3D counterparts, cf. Eq. (\ref{eq5}). However, numerical analysis
demonstrates that all such higher-order states, with $s\geq 1$, are
completely unstable, in the 2D and 3D systems alike.

\section{Conclusion}

This article aims to present a brief review of the broad area of 2D and 3D
solitons, both fundamental (zero-vorticity) and vortical ones. The
multidimensional solitons and solitary vortices find most important physical
realizations in BEC and nonlinear optics, the most essential problem being
the search for physically relevant settings which provide for the
stabilization of this states against the collapse and splitting, that tend
to destroy the fundamental and vortical solitons, respectively, in media
with the (most common) cubic self-attractive nonlinearity. The presentation
includes two particular topics, one well-established, \textit{viz}., the
stabilization of the 3D and 2D modes with vorticities $S=0$ and $1$ in a
trapping harmonic-oscillator potential (generally, anisotropic one), and the
most recent topic, namely, the creation of stable 2D and 3D solitons, which
mix terms with $S=0$ and $1$ (semi-vortices) or $S=0$ and $\pm 1$ (mixed
modes), in a two-component system which realizes the spin-orbit coupling in
BEC. In either case, the analysis reveals a drastic difference between the
2D and 3D settings. In the former case, the stabilization mechanism creates
GSs (ground states), that were missing without it. The total norm of the GSs
takes values below the threshold necessary for the onset of the critical
collapse, driven by the cubic attractive nonlinearity in the 2D space. In
the 3D settings, the supercritical collapse does not allow the creation of a
GS, but, nevertheless, an appropriate mechanism may create metastable
solitons, which are robust against small perturbations.

There are many directions for the development of the studies of
multidimensional solitons. As concerns the theory, a completely different
possibility is the use of spatially nonuniform \emph{self-repulsive}
nonlinearity, with the local strength growing from the center to periphery
at any rate faster than $r^{D}$, where $r$ is the distance from the center,
and $D$ the spatial dimension. This model makes it possible to create very
robust solitons of diverse types, including fundamental ones and solitary
vortices \cite{ICFO,further}, as well as sophisticated 3D states -- notably,
\textit{hopfions} (vortex tori with inner twist, which feature two
independent topological numbers). A challenging problem is finding still
more general physically relevant conditions for the creation of complex 3D
modes, such as the hopfions \cite{hopfion}, skyrmions (which, similar to
hopfions, carry two different topological charges) \cite%
{skyrmions,skyrmion-exper}, monopoles \cite{monopole}, linked vortex rings,
and others. In terms of the experiment, the entire area of multidimensional
solitons remains a challenging one, as very few experimental results have
been reported, thus far (the latest experimental findings are the creation
of (2+1)D optical spatial solitons in media with competing
focusing-defocusing cubic-quintic \cite{Cid1} and quintic-septimal \cite%
{Cid2} nonlinearities, as well as the making of a quasi-stable vortex
solitons in an optical medium where nonlinear losses are essential \cite%
{Cid3}).

\section*{Acknowledgments}

I would like to thank my collaborators in original works on various topics
related to multidimensional solitons (most of those works are cited in this
review): F. Kh. Abdullaev, C. B. de Ara\'{u}jo, B. B. Baizakov, V. Besse, O.
V. Borovkova, G. Boudebs, C. M. Brtka, W. B. Cardoso, R. Carretero-Gonz\'{a}%
lez, Z. Chen, J. Cuevas, R. Driben, N. Dror, P. Dyke, Z. Fan, D. J.
Frantzeskakis, A. Gammal, G. Gligori\'{c}, L. Had\v{z}ievski, R. G. Hulet,
X. Jiang, Y. V. Kartashov, P. G. Kevrekidis, V. V. Konotop, H. Leblond, B.
Li, Y. Li, V. E. Lobanov, D. Luo, A. Maluckov, D. Mazilu, T. Meier, D.
Mihalache, J. H. V. Nguyen, W. Pang, H. Pu, A. S. Reyna, H. Sakaguchi, L.
Salasnich, M. Salerno, E. Ya. Sherman, Ya. Shnir, H. Susanto, F. Toigo, L.
Torner, F. Wise, Y.-C. Zhang, and Z.-W. Zhou. I also thank Carlos
Pando-Lambruschini for the invitation to submit the paper to the special
issue of the European Physical Journal - Special Topics.

Parts of this work were supported by grants No. I-1024-2.7/2009 from the
German-Israel Foundation, and No. 2010239 from the Binational (US-Israel)
Science Foundation.

\end{document}